# A Novel Channel Boosted Residual CNN-Transformer with Regional-Boundary Learning for Breast Cancer Detection

Aamir Mehmood, Yue Hu,, Saddam Hussain Khan*

*Artificial Intelligence Lab, Department of Computer Systems Engineering, University of Engineering and Applied Sciences (UEAS), Swat 19060, Pakistan

Email: saddamhkhan@ueas.edu.pk

**Abstract**—Recent advancements in detecting tumors using deep learning on breast ultrasound images (BUSI) have demonstrated significant success. Deep CNNs and vision-transformers (ViTs) have demonstrated individually promising initial performance. However, challenges related to model complexity and contrast, texture, and tumor morphology variations introduce uncertainties that hinder the effectiveness of current methods. This study introduces a novel hybrid framework, CB-Res-RBCMT, combining customized residual CNNs and new ViT components for detailed BUSI cancer analysis. The proposed RBCMT uses stem convolution blocks with CNN Meet Transformer (CMT) blocks, followed by new Regional and boundary (RB) feature extraction operations for capturing contrast and morphological variations. Moreover, the CMT block incorporates global contextual interactions through multi-head attention, enhancing computational efficiency with a lightweight design. Additionally, the customized inverse residual and stem CNNs within the CMT effectively extract local texture information and handle vanishing gradients. Finally, the new channel-boosted (CB) strategy enriches the feature diversity of the limited dataset by combining the original RBCMT channels with transfer learning-based residual CNN-generated maps. These diverse channels are processed through a spatial attention block for optimal pixel selection, reducing redundancy and improving the discrimination of minor contrast and texture variations. The proposed CB-Res-RBCMT achieves an F1-score of 95.57%, accuracy of 95.63%, sensitivity of 96.42%, and precision of 94.79% on the standard harmonized stringent BUSI dataset, outperforming existing ViT and CNN methods. These results demonstrate the versatility of our integrated CNN-Transformer framework in capturing diverse features and delivering superior performance in BUSI cancer diagnosis.

## I. INTRODUCTION

BREAST cancer ranks high among causes of cancer deaths in women globally. Timely detection is crucial for effective treatment and control [1], [2]. Numerous imaging modalities, like ultrasound, MRI, histopathological, and CT, assist in identifying breast abnormalities. Breast ultrasound imaging (BUSI) is frequently used for tumor characterization, especially in patients with dense breasts [3]. Studies show that BUS surpasses mammography in diagnostic performance, offering benefits like noninvasiveness, portability, real-time imaging, affordability, and absence of ionizing radiation. Research has shown that breast screening can reduce breast cancer mortality by 30%–40% in women [4]. However, despite significant progress in breast screening, a considerable percentage of cancers, around 10% to 30%, are missed in diagnostic workup settings, and 30% are recalled for further evaluation of potential abnormalities [5].

The labor-intensive and subjective nature of the manual examination of these large and morphologically diverse slides remains challenging. Moreover, BUS is limited by illuminations, minor contrast, variable structure, and noise, often due to poor probe contact, improper pressure, or subcutaneous fat, which decreases noise. Therefore, there is an increasing interest in integrating deep learning (DL) techniques, especially convolutional neural networks (CNNs), into the pathological workflow to provide computer-aided diagnostic support. DL has shown promising results in automating pathological image analysis tasks, including cancer region detection, cancer subtyping, cancer grading, etc.

Advances in DL, particularly CNNs and vision-transformers (ViTs), have enhanced medical imaging [6]. CNNs learn dynamic feature representations of BUS images and achieve optimal detection performance [7]. Nonetheless, CNNs struggle with long-range dependencies, which can slow processing and limit generalization. ViTs, using self-attention mechanisms, effectively manage global dependencies [8]. CNNs face challenges in accurately capturing long-range dependencies, resulting in suboptimal detection of large-scale breast cancer. In comparison, conventional ViT methods encounter difficulties in local feature extraction and computational complexity [9]. This limitation stems from the inherent constraints of convolution, which restrict information aggregation to adjacent pixels and require multiple downsampling steps to achieve a sufficiently local receptive field and improve the computation. Research indicates that combining ViTs with CNNs improves tumor detection accuracy in BUS data.

However, ensemble methods have drawbacks, such as careful weight selection and equal importance assignment, which may be ineffective when models are correlated. Moreover, medical imaging challenges emerge due to limited datasets compared to extracted features, leading to the curse of dimensionality [10], [11], [12]. Transfer learning (TL) is utilized to mitigate overfitting and enhance model performance on unseen data [13]. Furthermore, datasets face limitations such as data scarcity, speckle noise, contrast, and morphology fluctuations with malignant and benign. Therefore, this study proposes an effective channel boosted (CB)-Residual CNN (Res)-regional and boundary CNN meet Transformer (RBCMT) technique for predicting breast malignant BUS images using diverse TL-based deep residual learning CNN and transformer combined with regional and boundary operations. Regional operations suppress noise and capture contrast, while boundary operations learn morphological variations. The novel channel-boosted (CB) strategy enhances feature diversity in the limited dataset by integrating original RBCMT channels with residual CNN-generated maps derived from transfer learning. Moreover, the residual learning captures texture information and handles vanishing gradients. The proposed CB-Res-RBCMT integrates an early-stage CNN block for efficient local feature extraction and reduced model complexity, and a two-stream network design to optimize information extraction for subsequent transformer processing. The proposed hybrid CNN-Transformer captures complimentary local and global information through residual learning, effectively addressing texture variation [14], [15]. This research contributes primarily to:

- A novel " CB-Res-RBCMT " hybrid technique merges

transformers and CNNs to analyze cancer in BUSI globally and locally. CB-Res-RBCMT incorporates three key elements: Regional and Boundary CNN Meet Transformers (RBCMT), residual learning CNNs, and Channel-Boosting (CB).

- The proposed RBCMT architecture features an abstract stem CNN and new CMT blocks, with a progressive arrangement of CNN layers incorporating regional and boundary (RB) feature extraction operations. Regional operations effectively mitigate noise and enhance contrast, whereas boundary operations adeptly capture morphological variations. The multi-head attention within the proposed CMT captures interactions by encapsulating each entity with global contextual information. Additionally, a light attention operation enhances computational efficiency for effective local information extraction.

- The CB-Res-RBCMT approach integrates RBCMT channels and TL-based Residual CNN blocks to enhance diverse feature space for target-level attention. Residual Learning-based CNN blocks capture essential local malignant features, while the new inverse residual block concept in proposed CMT blocks aids gradient propagation.

- Finally, a new pixel-attention (PA) mechanism optimizes pixel selection to highlight subtle discriminative patterns and minimize minor inter-class variations across cancerous and benign. The proposed CB-Res-RBCMT techniques are rigorously tested on a publicly available, harmonized, stringent BUSI dataset for accurate breast ultrasound analysis and benchmarked against leading state-of-the-art CNNs and ViTs.

The manuscript is designed as follows: Section 2 analyses related concepts, Section 3 presents the proposed breast tumor detection framework, and Section 4 describes the dataset, implementation, and performance metrics. Section 5 offers a detailed analysis of the results, including an ablation study and systematic comparisons. Section 6 concludes the manuscript and considers future research directions.

## II. Related Work

This section succinctly reviews recent automatic breast tumor diagnosis methods using BUSI and DL [16]. These methods leverage prior knowledge or apply identification algorithms to entire medical datasets to address diagnostic challenges. Typically, DL models are initially trained on natural data like ImageNet and tuned on specific medical datasets [17], [18], [19], [20]. Several DL methods have been reported to detect breast tumors as benign or malignant. The enhanced VGG16 with an attention mechanism improves the extraction of relevant features and distinguishes important pixels in ultrasound images using cosine loss functions and achieved 93% accuracy [21]. Luo et al. [22] developed a segmentation model to create a binary map of breast tumors in ultrasound images. Two parallel networks then processed the original and segmented images, and an attention-based feature aggregation model enhanced classification performance. However, its complexity and longer training time have significant limitations. Byra [23] introduced deep representation scaling layers in a TL approach, reducing trainable parameters and achieving 91.5% classification accuracy. Du et al. [24] adapted the EfficientDet method to detect breast masses as normal, benign, or malignant. Han et al. [25] used GoogLeNet on 7,408 ultrasound samples from 5,151 patients, applying data augmentation to achieve 90% accuracy. Qi et al. [26] reported two CNNs to detect breast tumors using activation maps from each model to guide the other, evaluated on breast ultrasound images. Hassanien et al. [27] presented a radiomics approach utilizing ultrasound sequences. They employed the ConvNext network for feature extraction and implemented a malignant tumor score.

Several studies have employed ensemble approaches to classify breast tumors in BUSIs. Tanaka et al. [7] combined outputs from VGG19 and ResNet152 models to classify malignant and benign tumors. Moon et al. [28] developed a CAD system employing a fusion approach that incorporated original BUSIs and applied an ensemble of CNN outputs. This method required experts to manually define ROIs and tumor contours, leading to increased time consumption and variable segmentation accuracy. A Mask-RCNN has been developed to generate segmentation outputs simultaneously, aiding in identifying malignant tumors by generating probabilities [29]. Ragab et al. [30] developed a classification framework using features from VGG16, VGG19, and SqueezeNet, optimized with swarm optimization. Misra et al. [31] used an ensemble approach, combining features from AlexNet and ResNet to detect malignant tumors in ultrasound and elastography images. While these methods achieved significant classification accuracy, they often lacked consideration for the reliability of detection techniques, which is crucial for clinical relevance.

Few studies have explored Transformer methods for diagnosing breast tumors using medical image datasets [32]. Gheflati et al. [8] used pre-trained ViT for detecting tumors using two BUS datasets to prevent overfitting and improve feature representation, achieving comparable results to existing CNNs. However, recent efforts to develop large end-to-end DL models demand high-performance resources and traditional ViT methods often struggle with capturing local features and computational complexities. The previous research work has the following limitations:

- Existing CNN models trained on natural images struggle with medical images due to distinct patterns and textures, while CNNs' focus on local features can reduce spatial correlation, limiting performance on complex patterns.

- Traditional ViT models suffer from high computational complexity, long training times, and inefficient local feature extraction, reducing practical deployment.

- DL models face vanishing gradient issues in deeper networks and are constrained by limited dataset diversity in BUSI studies, hindering learning efficiency and generalization.

In this regard, combining the strengths of CNNs and customized ViT techniques could enhance both global and local feature performance and clinical applicability.

## III. Methodology

The proposed framework integrates the novel CB-Res-RBCMT and existing ViT/CNNs for cancerous (malignant) tumor classification from non-cancerous (benign) using BUSIs. Data augmentation is employed as a preprocessing method to condense bias and improve generalization. The proposed tumor detection approach is compared with the evaluation of existing ViTs and hybrid CNN-ViT, and the existing CNNs. Figure 1 illustrates the overall breast tumor analysis workflow.

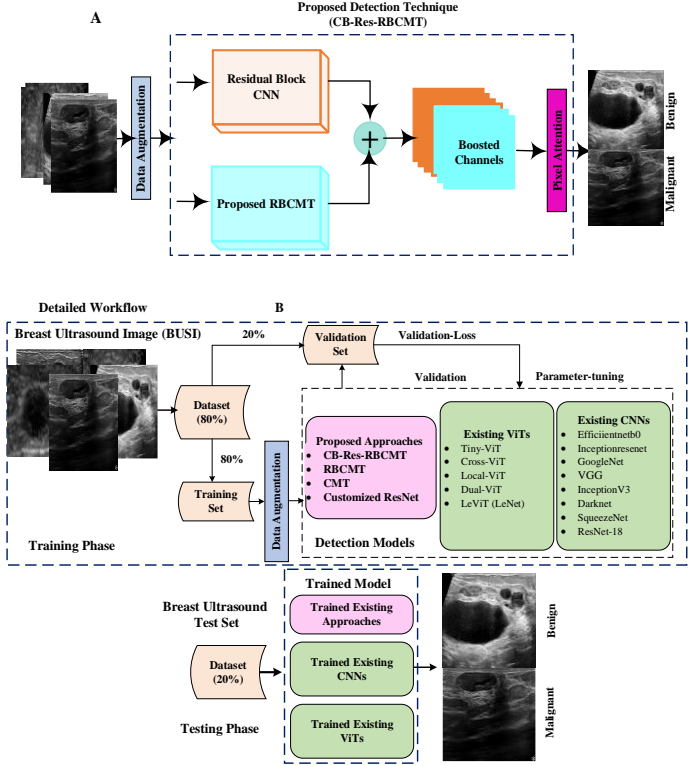

**Fig. 1:** (A) Brief and (B) detailed overview of the proposed Breast Cancer Analysis framework.

## A. Data Augmentation

Data augmentation is employed in the preprocessing phase to address class imbalance and overfitting. The dataset's imbalance leads to elevated false positives and impacts results. To mitigate this, data augmentation enhances the representation of underrepresented classes using techniques such as flipping, scaling, reflection, and shearing.

## B. The Proposed CB-Res-RBCMT Technique

The proposed CB-Res-RBCMT framework consists of the novel RBCMT architecture and TL CNN residual blocks for generating and concatenating auxiliary feature maps. The backbone design includes four stages, each with a parallel sequence of RBCMT and residual blocks. Inspired by the Hybrid ViT-CNN, which is prominent in medical image analysis, the RBCMT addresses the limitations of ViTs. ViTs segment images into linear, non-overlapping patches processed through encoder blocks using linear layers, which may not capture local and structural information effectively. In contrast, CNNs excel at capturing image-specific locality, translational equivariance, and correlated features within two-dimensional neighborhoods.

The RBCMT is designed to prioritize convolutional efficiency, particularly in the early stages of breast cancer image processing, for patching and tokenization. This focus enhances the model's ability to extract relevant features. TL-based residual learning generates diverse, rich, high-dimensional feature maps. The feature maps produced by the RBCMT and CNN residual blocks are combined to capture diverse and rich information. The FME-Residual-CMT Net also employs an attention module to focus on subtle discriminative patterns and intra-class contrast variations in breast tumors. Figure 2 illustrates the proposed RBCMT and the integration of residual learning within the CB-Res-RBCMT framework. Details of each module are explained below.

### 1) Proposed RBCMT

Our research aims to develop an integrated network that influences the strengths of CNNs and transformers. The proposed RBCMT architecture combines an initial stem CNN with customized CMT blocks, followed by sequential CNN layers that integrate both regional and boundary (RB) feature extraction processes. In the stem CNN block, images undergo initial processing, are divided into patches, and then global features are extracted using embedded patch tokens within the transformer blocks, as shown in Figure 2. The stem CNN block uses a 3×3 convolution (Conv_L), stride (2), and an output channel (64), followed by two additional 3×3 Conv_L layers and stride (1) to enhance fine-grained information extraction.

The proposed RBCMT architecture comprises four stages designed to generate feature maps at varying scales, which is crucial for diagnosis challenges. After each stage, Conv_L and activation are applied, followed by max and average pooling to enhance regional homogeneity and capture structural and boundary information (Equations 1-3). Moreover, regional operations robustly suppress noise and amplify contrast, while boundary operations proficiently model complex morphological variations.

The input image's feature dimension is increased to 64 in the initial convolution block and doubles after each pooling operation. This process downscales the intermediate feature resolution at the end of each stage, capturing invariance and improving robustness. The patch-embedding layer generates tokens of various dimensions after the initial stages. Ultimately, the proposed RBCMT produces four hierarchical channels with distinct resolutions, similar to conventional CNNs.

$$c_{k,l} = \sum_{a=1}^{x} \sum_{b=1}^{y} c_{k+a-1,l+y-1} f_{a,b} \qquad (1)$$

$$c^{max}_{k,l} = \max_{a=1,\dots,w,b=1,\dots,w} x_{k+a-1,l+b-1} \qquad (2)$$

$$c^{avg}_{k,l} = \frac{1}{w^2} \sum_{a=1}^{w} \sum_{b=1}^{w} x_{k+a-1,l+b-1} \qquad (3)$$

The Conv_L extracted feature map, represented by '$c$' with dimensions 'k x l', is characterized in Equation (1) by the kernels denoted as '$f$' with a size of 'a x b'. The output spans [1 to k-x+1, l-y+1]. The homogeneous and structural operation window size, denoted by '$w$' is applied to the convolved output ($c_{k,l}$) in Equations 2-3.

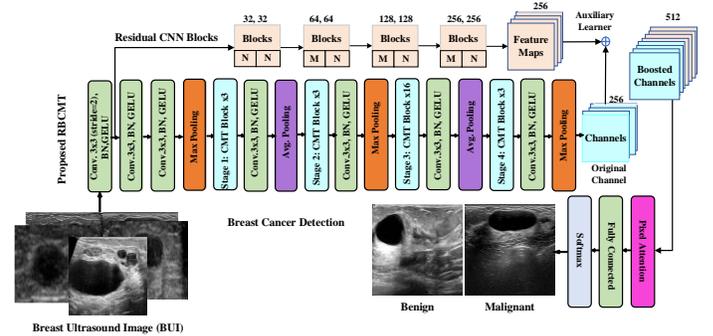

**Fig. 2:** The proposed CB-Res RBCMT technique, comprised of the proposed RBCMT integrating stem-CNN and proposed CMT followed by region and boundary operation, and TL-based residual M and N CNN blocks.

### i). Proposed CMT Block

The proposed CMT architecture, comprising four stages meticulously designed to generate channels of varying scales, is crucial for addressing prediction challenges. Each stage systematically stacks CMT blocks, facilitating feature transformation while preserving input resolution. Figure 3 visually illustrates the incorporation of CMT blocks in each stage. Prior to each stage, a patch embedding layer, incorporating 3×3 Conv_L and normalization (Norm) techniques, establishes a hierarchical representation within the proposed CMT. Additionally, the customized CMT enables the generation of multi-scale representations for downstream tasks, with various configurations and stridden Conv_L following the input, particularly advantageous for breast tumor analysis.

The CMT block adeptly captures both local and long-range dependencies, as detailed in Section 3.2. In line with the ViT concept, the proposed CMT block deviates from traditional MSA, introducing a lightweight (LMSA) block and replacing the MLP layer with a new Inverted Residual Feed-Forward Network (IRFFN). This includes two layer-normalization sub-layers followed by LMSA and IRFFN, respectively. These architectural refinements are pivotal in enhancing the ability to capture intricate dependencies and enrich feature representation for improved performance in complex tasks. Furthermore, to augment the network's representation capacity, a Local Perception Unit (LPU) is integrated into the CMT block, as depicted in Figure 3. Finally, the proposed CMT block is formulated with these three components as follows and mathematically expressed in Equations 4-6.

$$y_i = LPU(c_{i-1}) \tag{4}$$
$$z_i = LMHSA(LN(y_i)) + y_i \tag{5}$$
$$c_i = IRFFN(LN(y_i)) + z_i \tag{6}$$

The outcome of the $i^{th}$ block from LPU and LMHSA blocks is represented as $y_i$ and $z_i$. LN is implemented, and multiple CMT blocks are sequentially stacked in each stage to enable efficient feature transformation and aggregation.

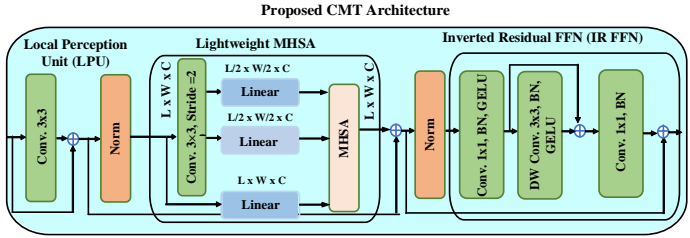

**Fig. 3:** The proposed CMT blocks featuring LPU, lightweight MHSA, and IRFFN.

### A. Local Perception Unit (LPU)

The LPU addresses challenges of invariance, particularly addressing rotation and shift data augmentation, thus preserving translation invariance in model outcomes [27]. In contrast to prior transformers utilizing absolute positional encoding, which assigns unique positional encoding to each patch and disrupts invariance [6], ViTs overlook local relations [38] and structural information [26] within patches, necessitating an improved approach. Therefore, the LPU is introduced to extract local information and overcome these limitations. The input, denoted as '$c$', has sizes L, W, and C, where L×W and d represent input and feature dimension, respectively (Equation 7).

$$LPU(c) = Conv\_L(c) + c \tag{7}$$

### B. Light Weight Multi-Head Self-Attention (LMHSA)

The Multi-Head Self-Attention (MSA) module addresses the limitations of a single-head SA module, which may focus on a limited number of positions, potentially neglecting other crucial ones. SA is a critical component of ViTs, known for its explicit representation of relationships among entities within a sequence. Various ViT architectures have emerged, modifying the SA module to enhance effectiveness. Some models incorporate dense global attention mechanisms, while others leverage sparse attention mechanisms to capture global-level dependencies in spatially uninformative images. MSA addresses this constraint by parallel stacking of SA blocks to enhance the effectiveness of the SA layer. This mechanism encapsulates each entity with global contextual information and captures their interactions [30]. In the initial SA block, the input map '$x$' is linearly transformed to produce query ($q$), key ($k$), and value ($v$) matrices. This involves assigning distinct depiction subspaces ($q, k,$ and $v$) to the attention layers, enabling MSA to learn diverse and intricate interactions among sequence elements (Equations (8-10)), as detailed in Equation 3.

$$A(q, k, v) = \sigma\left(\frac{q.k^T}{\sqrt{d_k}}\right) \tag{8}$$
$$k' = Conv\_L(k) \tag{9}$$
$$v' = Conv\_L(v) \tag{10}$$

Additionally, a 3×3 depth-wise Conv_L with a stride of 2 is used to decrease the spatial dimensions of k and v before the attention operation, enhancing computational efficiency and enabling efficient extraction of local information. The initial stages of the transformer are substituted with convolution blocks, leading to downsized feature maps of various dimensions. These maps are then token-embedded and processed by transformer blocks for feature extraction. A 4×4 window patch size is employed in the attention mechanism to improve processing granularity. Furthermore, a relative position bias **B** is integrated into the SA block, defining the corresponding lightweight attention (LA) as follows:

$$LA(q, k', v') = \sigma\left(\frac{q.k'^T}{\sqrt{d_k}} + B\right)v' \tag{11}$$

The attention mechanism, symbolized by 'LA' and 'σ', involves activation. Here, $q$, $v$, $k^T$ represent the query, value, and transposed key matrices, respectively (Equation 11). Additionally, $\sqrt{d_k}$ serves as a scaling parameter, with '$d_k$' representing the dimension of the key matrix.

The MSA consists of multiple SA blocks, each featuring learnable weight matrices for $q$, $k$, and $v$ subspaces. These blocks produce outcomes that are combined and transformed using the trainable parameter $w^o$, projecting them into the output space. Furthermore, the proposed CMT demonstrates significant adaptability for fine-tuning across various downstream detection challenges. The LMHSA module utilizes "heads," employing LA functions to generate a sequence of dimension n × d/h. These sequences from multiple heads are then merged into unified n (resolution/number of patches) × d dimensions of $q, k, v$ features (Equations 12-13), where 'h' represents the head and 'cat' is a concatenation operation, mathematically expressed as follows:

$$LMHSA(q, k', v') = cat(h_1, h_1, \dots, h_h).w^o \tag{12}$$
$$h_i = LA(q_i, k'_i, v'_i), \text{ where i=1,2, ..., h} \tag{13}$$

### C. Inverted Residual Feed-Forward Network

The proposed Inverted Residual Feed-Forward Network (IRFFN) resembles the residual block, optimizing performance by adjusting the location of the shortcut connection while incorporating an expansion layer and convolution. Typically, the FFN consists of two linear layers separated by GELU activation, extracting and integrating complex features (Equation 14) [33]. The FFN is utilized in each encoder block following the SA block. We introduced 3×3 and point-wise (1x1) Conv_L in the transformer FFN block to reduce parameters and computational complexity while maintaining performance. Moreover, IRFFN introduces a skip connection inverted residual block and replaces the activation function GELU and normalization. The introduction of a shortcut follows the principles of classic residual networks, promoting gradient propagation across layers [34] (Equations 15-16). Additionally, utilizing Conv_L captures efficient local information extraction. Equation 14 represents the non-linear activation function GELU, denoted by $\sigma_g$. Here, $w_1$ and $w_2$ represent weights, while b1 and b2 correspond to biases.

$$FFN(c) = (b_2 + w_2 * \sigma_g(b_1 + w_2 * c)) \tag{14}$$
$$IRFFN(c) = Conv(F(Conv(c))) \tag{15}$$
$$F(c) = Conv\_L(c) + c \tag{16}$$

#### 2) CNN Residual Block

We implemented a refined stacking technique that combines TL-based CNN residual learning with four M and N blocks for systematic learning enhancement (Figure 4). PWC, utilizing a $1 \times 1$ kernel size in the M block, promotes inter-feature map communication and linearly projects Conv_L feature maps into distinct output dimensions. The Conv_L kernel size is set to $3 \times 3$ to achieve a local receptive field in both M and N blocks (Equation 1). Strategically concatenating these blocks with CMT in the final stage allows effective exploration of diverse feature spaces. This systematic arrangement, comprising four sequential residual blocks, facilitates the acquisition of a wide range of essential features. Channel numbers progressively increase from 64 to 256 to strengthen the learning process, ensuring meticulous and refined learning for improved outcomes.

The residual block employs TL to generate additional feature maps, thereby diversifying features. These added channels, derived from TL-based deep CNNs, dynamically capture subtle variations in representation and texture within MRI images of breast tumors. Leveraging residual features enables the discernment of intricate features essential for distinguishing contrast and texture variations of tumors in BUSIs, facilitated by the impactful deep CNN rooted in FME.

$$y = T(x, \{w_i\}) + x \qquad (17)$$
$$y = T(x, \{w_i\}) + w_s x \qquad (18)$$

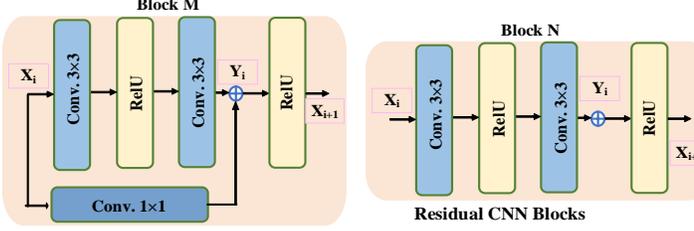

Fig. 4: Residual CNN Block.

### 3) Feature Map Enhancement

FME enhances the learning capabilities of the proposed RBCMT technique through TL-based residual learning, allowing auxiliary learners to discern diverse and intricate patterns within images. The CB-Res-RBCMT technique employs three heterogeneous concepts, incorporating the RBCMT, residual learning, and FME by concatenating diverse channels (equation 19) to capture multi-level variations. CNN and CMT-based channels specialize in capturing local-level diversity and global target-level features in image patterns, respectively. Additionally, residual learning CNN integrated with CB (CB-RBCMT) captures texture variation and demonstrates notable performance for breast tumor analysis. Furthermore, FME-residual learning retains class-specific information at both the channel and spatial levels. The process of channel concatenation is depicted in Equation 19, denoted by the symbol ||, and Figures 1-2.

$$c_{Boosted} = b(x_{Res} || c_{RBCMT}) \qquad (19)$$

Equation (19) represents the channels of the RBCMT and Residual block $c_{RBCMT}$ and $x_{Res}$, respectively. Furthermore, the additional feature-maps from block R obtained through TL are denoted as $x_{Res}$. The concatenation operation $b(.)$ enhances the feature maps.

### 4) Pixel-Attention (PA)-based Approaches

We employed pixel attention (PA)-based block to capture class-specific features at spatial levels (Figure 5). The boosted maps from various learners are amalgamated and weighted through the attention concept (Equations 20-22). This allows the network to emphasize the most relevant pixels while disregarding redundant information [34]. The notation $\oplus$ denotes element-wise addition and multiplication, illustrating diverse boosted channel and spatial PA evaluation. The boosted feature map, denoted as $x_{Boosted}\cdot$, undergoes refinement through element-wise addition between various pixel-weighted activations and the input, resulting in computing $W_{pixel}$. Boosted maps are then element-wise multiplied with spatial attention, yielding the final refined output feature map, $x_{SA\_out}$. Finally, the sequentially weighted output from the SA block is provided with fully connected layers, represented mathematically in Equations (23-24). This process enhances the effective class-specific feature map and discriminative pixel contribution and improves model performance.

$$c_{SA\_out} = W_{pixel} \cdot c_{Boosted} \qquad (20)$$
$$x_{relu} = \sigma_1(W_c c_{Boosted} + W_{SA} SA_{m,n} + b_{SA}) \qquad (21)$$
$$W_{pixel} = \sigma_2(f(x_{relu}) + b_f) \qquad (22)$$
$$c = \sum_a^A \sum_b^B v_a \cdot c_{SA\_out} \qquad (23)$$
$$\sigma(d) = \frac{e^{c_i}}{\sum_{l=1}^d e^{c_d}} \qquad (24)$$

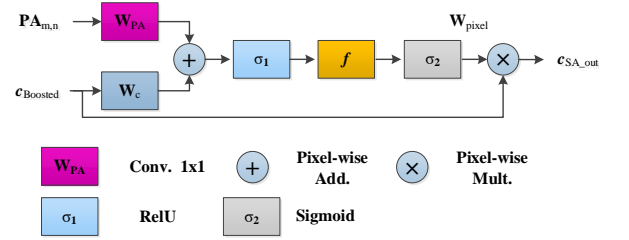

Fig. 5: Pixel Attention Block Designing.

Equation (20) represents boosted-map ($c_{Boosted}$) and $W_{pixel}$ as the weighted-pixel within the interval [0, 1]. The resulting output, denoted as $c_{SA\_out}$, highlights the cancerous patterns while mitigating the extraneous characteristics. Equations (21) and (22) elaborate on the activation $\sigma_1$ and $\sigma_2$, biases $b_{PA}$ and $b_f$, and transformations $W_c$, $W_{PA}$, $f$. The number of neurons and activation (softmax) in Equations (23-24) is expressed by $v_a$. and $\sigma$.

### C. Implementation of CNNs and ViT Techniques

Our study integrates contemporary ViTs and CNNs models to conduct a comparative analysis. To precisely delineate tumors in BUSIs, we utilize diverse datasets with various deep CNssNs and ViTs [41]. Various CNN and ViT models, including VGG-16/19, ResNet-50, ShuffleNet, Xception, Vit Tiny ViT, Cross ViT, LocalViT, and HVT models among others, are employed for cancer analysis [37]. CNNs have proven effective in detecting cancerous tumor images in the medical domain [19]. These deep CNNs, featuring diverse depths and network designs, are specifically adapted for analyzing various tumor types (benign and malignant). ViT models demonstrate heightened accuracy, even on smaller datasets such as CIFAR10. To enhance local feature modeling in ViTs approaches like LocalViT and LeViT leverage depthwise convolutions and adaptive position encoding. LeViT (LeNet) incorporates convolutional operations to capture spatial and low-level information.

## IV. EXPERIMENTAL SETUP

### A. Dataset

The research employed DL classification techniques on a harmonized stringent dataset formed by merging distinct BUS datasets: BUSI and UDIAT, and Baheya Hospital comprising an integrated dataset of 2120 images. The effective integration of diverse imaging characteristics enhances the proposed model's robustness and generalizability in breast ultrasound analysis. The UDIAT BUSI dataset, sourced from the UDIAT diagnostic center of Parc Tauli Corporation in Sabadell, Spain [35], and resized resolution of $224 \times 224 \times 3$ pixels. The dataset was divided into training, validation, and test groups with a distribution shown in Table 1, ensuring a proportional representation of malignant and benign BUSI. Experienced medical professionals collected the data using Siemens Acuson, GE, and ATL HDI scanners. One professional manually identified regions of interest (ROIs) to demarcate breast mass regions.

Malignant images have been confirmed via biopsy by an expert pathologist, while benign images have been confirmed through biopsy or a two-year clinical history. Several BUS images of malignant and benign from the dataset are shown in Fig. 1. The proposed analysis technique has been evaluated using three publicly available BUSI image datasets, solely for testing. The OASBUD dataset contained 163 ultrasound images, including 110 benign and 53 malignant masses. All datasets underwent preprocessing, including the removal of scanner annotations, resizing images to 224 x 224 via bicubic interpolation, and the application of a 3 x 3 median filter.

**TABLE I.** DETAILS OF THE BUS DATASET PROVIDE US WITH IMAGES OF DIFFERENT DISTRIBUTIONS.

| Distributions | **Benign** | **Malignant** |
|---|---|---|
| **Training (70%)** | 686 | 668 |
| **Validation (10%)** | 176 | 166 |
| **Testing (20%)** | 214 | 210 |
| **Total (100%)** | 1076 | 1044 |

### B. Implementation and Experimental Settings

The proposed CB-Res-RBCMT and existing ViTs/CNNs models employed the Adam optimizer during training, learning rate ($10^{-3}$), and decay of 85% every 20 epochs while maintaining a constant weight-decay of 0.04. To address the class imbalance, the cross-entropy was utilized to evaluate the classification loss, batch size (16), and dropout (0.3) in the output layer. All experiments were conducted using Matlab 2024a on hardware featuring an Intel Core i7-10G, 32 GB of RAM, and a GPU NVIDIA-GeForce-GTX series (32 GB memory).

Utilizing data from diverse sources in the BUSI stringent dataset, performance evaluation was conducted through hold-out cross-validation, allocating 10% non-overlapping validation data in each iteration. The results across these validation sets were aggregated for performance assessment. Model performance was evaluated using various metrics (Equation 25-28), including F1-score, accuracy (Acc), sensitivity (Sen)/recall, precision (Pre), ROC/PR curve, and corresponding AUC. Equation (29) outlines the computation of the Standard Error (S.E.) for Sen within a 95% Confidence Interval (CI), aiming to enhance the True Positive (TP) rate and minimize False Negatives (FNs) in cancerous tumor analysis [36]. The z-value of 1.96 represents the S.E. within the 95% CI.

$$\text{Acc} = \frac{\text{TP+TN}}{\text{Total}} \times 100 \quad (25)$$

$$\text{Sen} = \frac{\text{TP}}{\text{TP+FN}} \times 100 \quad (26)$$

$$\text{Pre} = \frac{TP}{TP+FP} \times 100 \quad (27)$$

$$F-\text{score} = 2 \, x \, \frac{\text{Pre} \times \text{Sen}}{\text{Pre} + \text{Sen}} \quad (28)$$

$$C.I = z\sqrt{\frac{error(1-error)}{\text{Total Samples}}} \quad (29)$$

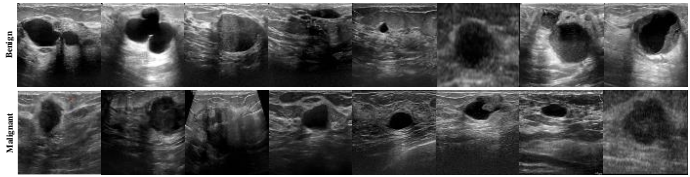

Fig. 6: Breast Ultrasound Image data comprises Benign and Malignant.

## V    RESULT AND DISCUSSIONS

This section outlines the evaluation details of the experiments and subsequently conducts ablation studies to evaluate the strength of the developed integrated RBCMT and CB-Res-RBCMT. Additionally, the performance of CB-Res-RBCMT is compared to previous works and existing ViTs/CNNs and two-stream (ViT-CNN) on the benchmark BUSI dataset in terms of ACC, Sen, Pre, F1-score, and AUCs, as depicted in Table 2 and Figures 7-11. The proposed techniques for binary classification of BUS images into two tumor categories: benign (non-cancerous) and malignant (cancerous). The proposed RBCMT outperformed existing ViTs/CNNs and achieved Acc (94.65%), Sen (93.93%), Pre (95.37%), and F1-score (94.57%), summarized in Table 2. The integration of CNN and transformer in the backbone improved performance metrics on the BUSI dataset. Moreover, the proposed CB-Res-RBCMT outperformed the customized RBCMT and achieved Acc (95.63%), Sen (94.79%), Pre (96.42%), and F1-score (95.57%), along with notable PR-AUC (0.9807) and ROC-AUC (0.9845). The results affirmed the effectiveness of the FME two-stream structure, RBCMT, and residual learning, and the early-stage CNN in enhancing feature encoding and generalization performance. Finally, the proposed CB-Res-RBCMT improves the respective classes' detection rate by enhancing TP while reducing the FN (8) and FP (6) of the proposed RBCMT and existing ViT/CNNs. We assessed the complexity of our lightweight and skip connection customized RBCMT model for accurate cancerous tumor detection from non-cancerous tissue using BUSI. Notably, lightweight RBCMT models required fewer computing resources and training time (approximately 12 min/epoch) than other ViT models and achieved optimal performance with a streamlined process. Moreover, this ensures faster and smoother convergence towards the solution and efficient training, suitable for resource-limited hardware, enhancing breast cancerous tumor detection accuracy. However, the good performance in the existing hybrid (CNN-ViT) fluctuates during convergence to reach the optimal solution.

**TABLE II.** PERFORMANCE ANALYSIS OF THE PROPOSED TECHNIQUES AND EXISTING CNNs/ViTs.

| Model | Accuracy | Sensitivity | Precision | F1-Score± S.E. |
|---|---|---|---|---|
| **Existing CNNs** | | | | |
| Efficiientnetb0 | 83.02 | 84.11 | 82.57 | 83.33 |
| Inceptionresenetv2 | 83.96 | 80.37 | 86.87 | 83.50 |
| GoogleNet | 84.43 | 87.85 | 82.46 | 85.07 |
| InceptionV3 | 86.32 | 86.92 | 86.11 | 86.51 |
| Darknet-53 | 87.26 | 85.05 | 89.22 | 87.08 |
| SqueezeNet | 86.79 | 89.72 | 84.96 | 87.27 |
| VGG-16 | 87.26 | 92.52 | 83.90 | 88.00 |
| DarkNet-19 | 88.21 | 91.59 | 85.96 | 88.69 |
| ResNet-18 | 89.62 | 90.65 | 88.99 | 89.81 |
| **Existing ViTs** | | | | |
| ViT | 87.26 | 86.92 | 87.74 | 87.32 |
| Cross ViT | 87.74 | 91.59 | 85.22 | 88.29 |
| LocalViT | 88.21 | 89.72 | 87.27 | 88.48 |
| Dual-ViT | 89.15 | 86.92 | 91.18 | 89.00 |
| (Hybrid ViT) Levit | 91.04 | 92.52 | 90.00 | 91.24 |
| **Proposed Setup** | | | | |
| **Customized ResNet** | 91.51 | 91.30 | 91.88 | 91.59 |
| **Proposed CMT** | 93.40 | 95.33 | 91.89 | 93.58 |
| **Proposed RBCMT** | 94.65 | 93.93 | 95.37 | 94.57 |
| **Proposed CB-Res-RBCMT** | 95.63 | 94.79 | 96.42 | 95.57 |

**TABLE III.** PERFORMANCE COMPARISON OF THE RECENT METHOD UTILIZED ON THE BUSI CANCEROUS TUMOR DATASET.

| Methods | Accuracy | Sensitivity | Precision | F1-Score |
|---|---|---|---|---|
| VGG16 [37] | 88.54 | 88.54 | 88.54 | 88.43 |
| VGG19 [37] | 88.28 | 88.28 | 88.20 | 88.07 |
| ResNet-50 [37] | 88.85 | 88.85 | 89.95 | 87.67 |
| Ensemble(VGG19/ ResNet152) [7] | --- | 90.90 | --- | --- |
| CNNs Ensemble [28] | 90.77 | 89.00 | 92.93 | 92.90 |
| DRS [23] | 91.50 | 90.40 | --- | --- |
| VGG19 [37] | 87.80 | 83.80 | 80.80 | 83.80 |
| MIB-Net [37] | 92.97 | 92.97 | 93.21 | 92.85 |

| | | | |
|---|---|---|---|
| Atn-VGG16 [21] | 93.00 | 96.00 | 92.00 | 94.00 |
| ViT_B_16 [38] | --- | 85.60 | 65.10 | 73.90 |
| ViT_S_16 [38] | --- | 66.50 | 71.40 | 68.80 |
| ViT_L_16 [38] | --- | 67.00 | 66.30 | 66.60 |
| BTS-ST (SwinT) [39] | --- | 93.30 | 79.20 | 85.60 |
| R+ViT/16 [8] | 85.70 | --- | --- | --- |
| ViT/16 [8] | 85.00 | --- | --- | --- |
| (CNN_ViT) Hybrid-MT-ESTAN [40] | 82.70 | 86.40 | --- | 86.00 |
| **UDIAT Dataset** | | | | |
| DAN [41] | 79.24 | 86.71 | 88.56 | 88.00 |
| DDC [41] | 77.65 | 89.57 | 86.33 | 88.00 |
| MADA [41] | 79.89 | 87.36 | 88.21 | 87.00 |
| DAAN-18 [41] | 80.78 | 88.13 | 87.67 | 87.00 |
| DAAN-50 [41] | 81.55 | 90.28 | 88.91 | 89.00 |
| MK-DAAN [41] | 83.16 | 94.31 | 89.26 | 91.00 |

### A. Performance Comparison with Existing ViTs/CNNs

This section compares the proposed CB-Res-RBCMT technique for cancerous tumor detection from benign with other leading models assessed on the BUSI datasets. Various CNN and ViT models, including VGG-16/19, ResNet-50, ShuffleNet, Xception, Tiny ViT, Cross ViT, LocalViT, Dual ViT, and HVT models among others, are employed for breast cancer analysis. Table 2 illustrates our model's performance in multi-classification tasks compared to previous advanced DL methods on the same dataset. The proposed technique demonstrates substantial improvements over conventional local correlated feature learning CNNs, with enhancements ranging from 6.01% to 12.61% for Acc, 2.27% to 14.42% for Sen, 7.2% to 13.96% for Pre, and 5.76% to 12.24% for F1-score (Figure 8). Additionally, it outperforms existing global receptive learning ViTs with improvements in Acc (4.59% -8.37%), Sen (2.27%-7.87%), Pre (5.24%-11.20%), and F1-score (4.33% to 8.25%), along with a notable increase of 3.62% and 5.11% in PR- and ROC-AUC, respectively. Moreover, a comparative analysis of the most recent previous techniques employed on cancer BUSI is detailed in Table 3. Recent transformer-based Dual-ViT models in Table 2 exhibit relatively better performance compared to existing CNN models. Existing Dual-ViT also outperforms other ViT models across most metrics, particularly excelling in Acc and Pre (91.04% and 92.52%), and ROC-AUC at 0.9639. In contrast, the LeViT method, rooted in a backbone network based on LeNet CNN and ViT methods, incurs a relatively lower computational overhead while achieving good performance compared to existing CNNs and ViT.

### B. Ablation Study of the Proposed Technique

In CB-Res-RBCMT, we optimized convolutional and transformer blocks using novel RBCMT and TL-based residual learning CNN. To assess their impact on performance and efficiency, we conducted an ablation study with varied block configurations, detailed in Table 2. Existing CNNs lack global receptive fields, while ViT models lack efficient feature extraction. Therefore, the integration of the hybrid LeViT (CNN-ViTs) facilitated the extraction of both local and global learning capacities, further enhancing performance but facing computational complexity. However, the absence of the early-stage CNN method in the backbone slightly degraded the model's generalization ability on unseen data. To address this, we substituted them with standard STEM-CNN blocks and customized lightweight transformer blocks in a hybrid approach. Consequently, the proposed lightweight CMT technique showed notable performance gains over the hybrid CNN-ViTs (LeViT), with improvements in Acc by 2.36%, Sen by 2.81.57%, Pre by 1.89%, and F1-score by 2.34%. Furthermore, the incorporation of systematic regional and boundary (RB) feature extraction operations with customized CMT using the BUSI dataset enhanced performance from Acc (93.40% to 94.65%), Pre (91.89-95.37%), and F1-score (93.58-94.57%). This demonstrated improved classification performance, with gains of Acc (5.50-7.40%), Pre (4.19-10.15%), and F1-score (5.57-7.25%) over transformer-based models. Finally, the proposed CB-Res-RBCMT incorporates lightweight techniques, a two-stream structure, and residual learning to simplify the backbone network and counter overfitting. Boosting the proposed RBCMT feature maps with residual learning CNN surpasses all existing ViTs and CNNs techniques, as evidenced by performance metrics, ROC-PR curves, and PCA-based analysis (Tables 2-3 and Figures (7-11)).

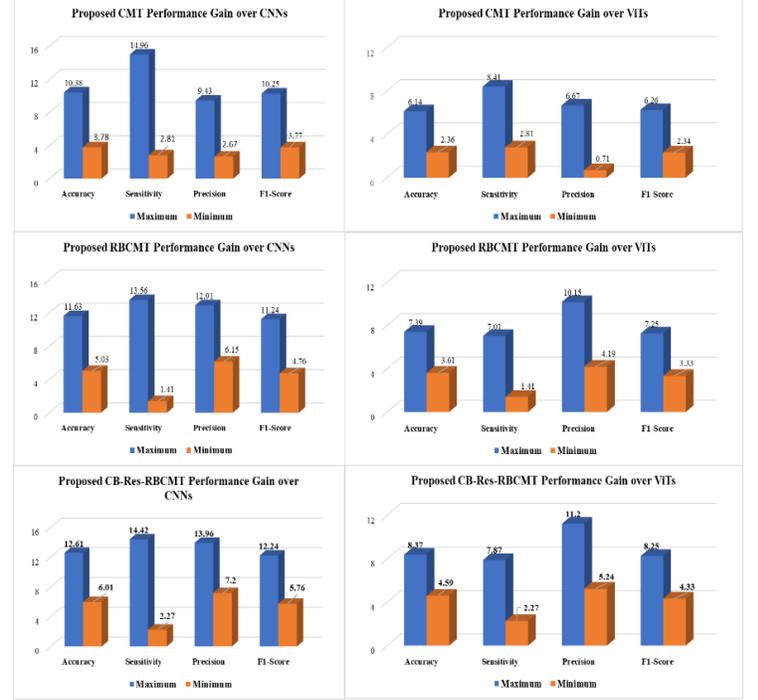

Fig. 8: The proposed CB-Res-RBCMT performance gain over existing techniques.

### C. PR/ROC curves

Detection rate curves quantitatively assess the discrimination ability of CB-Res-RBCMT across various threshold setups, evaluating its generalization between malignant and benign tumors. PR/ROC curves provide instant analysis of accurate prediction rates for breast cancerous (malignant) with non-cancerous (benign) (Figure 9). These curves compare predicted probabilities to ground-truth labels, with method performance summarized in Table 2 at 20% label availability. CB-Res-RBCMT demonstrates superior performance with a PR-AUC of 0.9807 and an AUC-ROC of 0.9845, outperforming all other active learning methods. Additionally, CB-Res-RBCMT shows improvements in AUC-PR ranging approximately from 3% to 20%, and gains in AUC-ROC of 2.5% to 9%. In Figure 9, our CB-Res-RBCMT exhibits the best ROC curves and the highest AUC among the compared CNNs/ViTs methods. Superior ROC performance is indicated by the deep purple curve and points closer to the top-left corner, while the purple curve represents better PR performance and points nearer to the top-right corner. The purple highlights indicate the proposed technique concentrates on identifying the malignant.

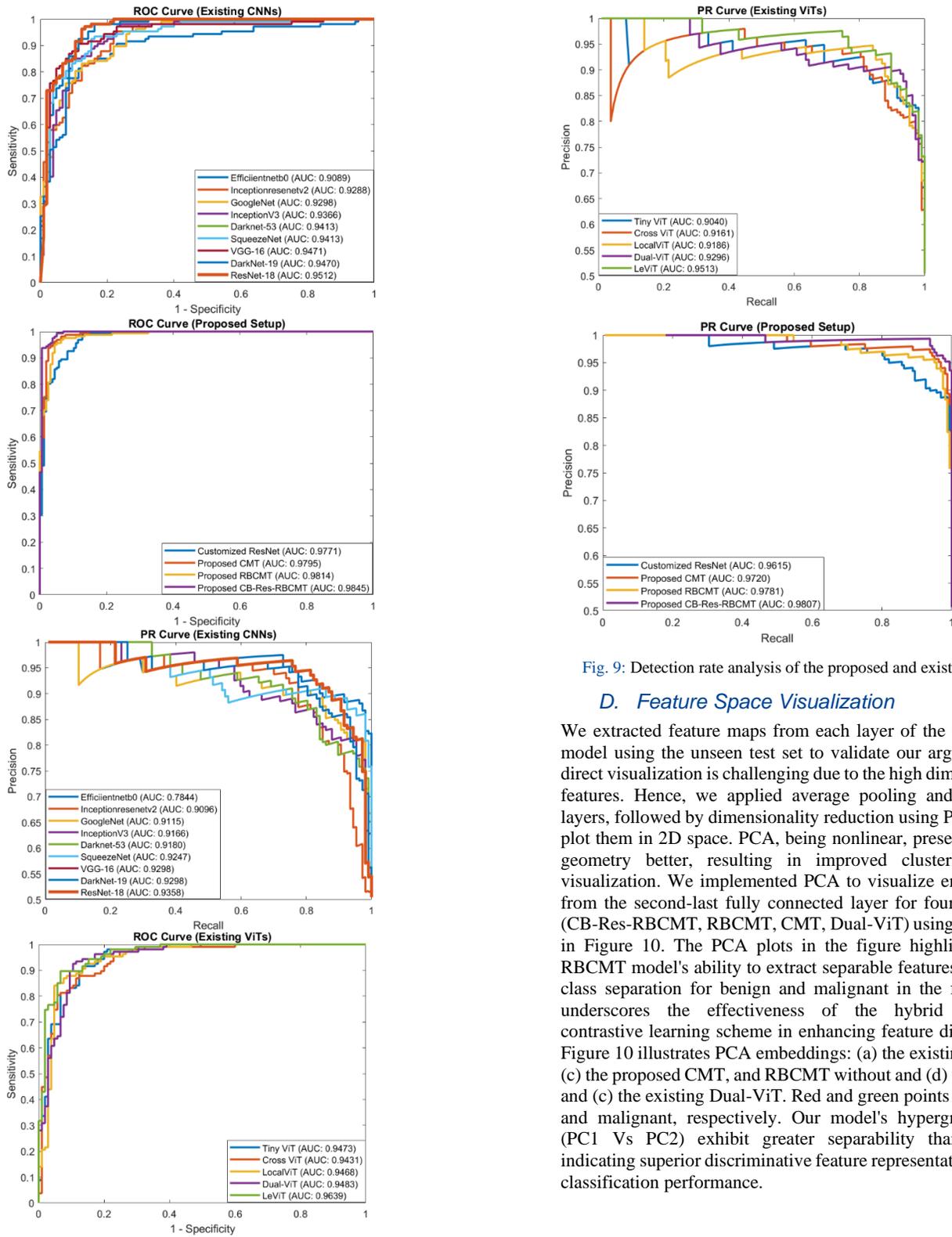

**Fig. 9:** Detection rate analysis of the proposed and existing CNNs/ViTs.

### D. Feature Space Visualization

We extracted feature maps from each layer of the CB-Res-RBCMT model using the unseen test set to validate our argument. However, direct visualization is challenging due to the high dimensionality of the features. Hence, we applied average pooling and fully connected layers, followed by dimensionality reduction using PCA techniques to plot them in 2D space. PCA, being nonlinear, preserves local feature geometry better, resulting in improved clustering quality and visualization. We implemented PCA to visualize embedded features from the second-last fully connected layer for four typical methods (CB-Res-RBCMT, RBCMT, CMT, Dual-ViT) using the BUSI dataset in Figure 10. The PCA plots in the figure highlight the CB-Res-RBCMT model's ability to extract separable features, leading to clear class separation for benign and malignant in the final layers. This underscores the effectiveness of the hybrid RBCMT-guided contrastive learning scheme in enhancing feature discriminativeness. Figure 10 illustrates PCA embeddings: (a) the existing Dual-ViT, (b)-(c) the proposed CMT, and RBCMT without and (d) with the CB term, and (c) the existing Dual-ViT. Red and green points represent benign, and malignant, respectively. Our model's hypergraph embeddings (PC1 Vs PC2) exhibit greater separability than other models, indicating superior discriminative feature representation and enhanced classification performance.

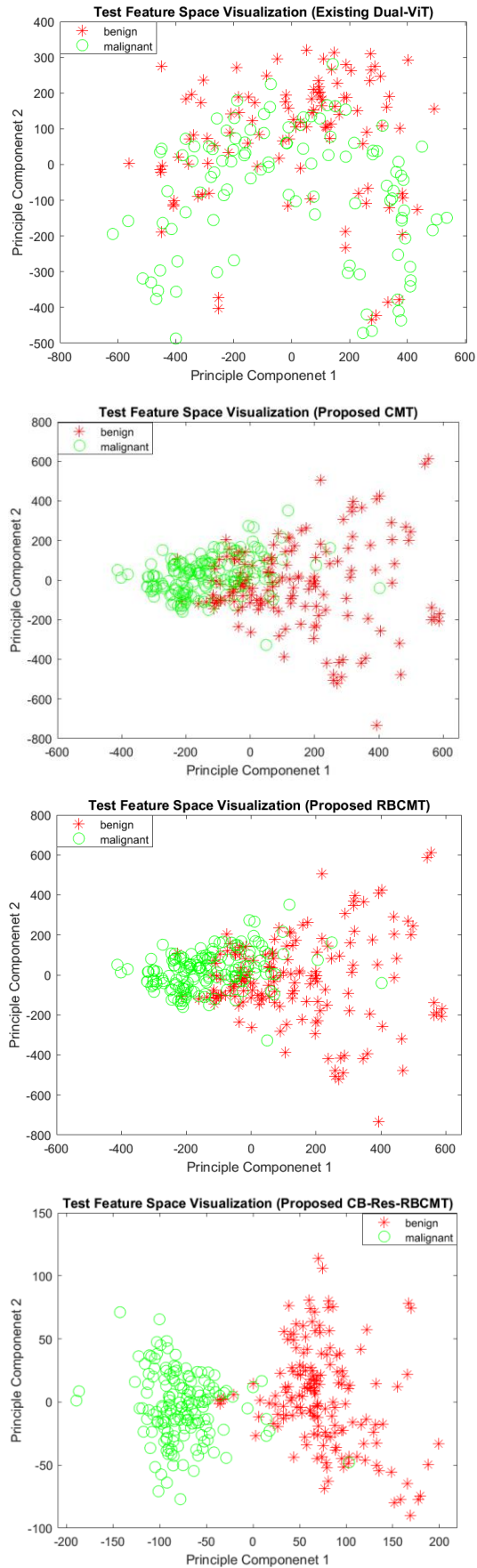

**Fig. 10:** Decision feature Visualization of the proposed setup on test samples.

### E. The Proposed Technique's Significance

- The proposed CB-Res-RBCMT Net integrates three heterogeneous architectures: RBCMT, residual learning, and CB, by combining diverse channels to capture multi-level variations.

- Our RBCMT method, inspired by the integrated approach of CNNs and transformer models, captures global and locally correlated features through stacked CMT blocks. Local diversity is emphasized by CNN-based channels, while CMT-based maps highlight global target-level features. Additionally, heterogeneous and structural operations address minor contrast and morphological information for inter-class variation and enhanced robustness.

- We introduced convolution blocks and lightweight transformer block versions to improve efficiency, reducing computational complexity and model capacity. Furthermore, the LPU maintains invariance for improved robustness, MSA computes global contextual interactions, and convolutional inverse residual connection captures local texture information within the CMT block.

- TL-based FME-residual learning CNN retains local and spatial levels information, effectively capturing cancerous texture variation. TL and data augmentation strategies are utilized to overcome limited and imbalanced data challenges.

- A novel pixel attention-based block captures class-specific features at spatial levels from boosted channels of various learners. This network highlights relevant pixels while disregarding redundant information, enhancing the effective class-specific feature map and discriminative pixel contribution.

- Extensive evaluations of the CB-Res-RBCMT technique on a harmonized BUSI dataset confirmed its superior performance, surpassing established CNN and ViT models in breast ultrasound analysis through rigorous benchmarking.

## VI. Conclusion

This study introduces a novel integrated approach, merging STEM and residual CNNs with a transformer technique for automated malignant tumor detection from benign using BUSI. Breast tumor images present challenges due to their complexity, including subtle contrasts, shadow, noise, morphological variations, and texture differences between benign and malignant classes. To address these challenges, a three-stream integrated CB-Res-RBCMT approach combines CNNs and transformer methods with TL-based residual FME learning to enhance feature extraction capabilities. The proposed integrated approach enhances consistency and effectively segregates cancerous (malignant) tumors from non-cancerous (benign) ones, addressing the image's complex illumination, texture, and morphology variability. The proposed technique outperforms existing hybrid LeNet CNN-ViTs, improving precision by 6.42%, accuracy by 4.59%, F1-score by 4.33%, and sensitivity by 2.27% for the harmonized stringent BUSI datasets. However, hybrid CNN and ViTs are computationally complex and face vanishing gradient issues. The proposed RBCMT integrates STEM convolution, lightweight transformer blocks, and downsized morphological operations to improve efficiency and reduce complexity. The architecture strategically incorporates an initial convolutional block for patch embedding and tokenization, processed by transformer blocks for comprehensive global feature extraction. Lightweight techniques reduce computational overhead while maintaining performance, demonstrating superior efficiency. Additionally, residual learning-based transformers and TL CNNs capture global and local texture variations, mitigating vanishing gradients. Novel spatial attention facilitates optimal pixel selection to enhance discrimination of minor contrast and inter-class pattern variation in stringent BUSI. Future work may incorporate multi-modal data collection or devising data augmentation algorithms like GANs may further improve model performance. Moreover, research will extend this method to estimate tumor region through segmentation and malignancy scores for liver and thyroid cancer, etc. in medical images.

## References

[1] L. W.- Sensors and undefined 2017, "Early diagnosis of breast cancer," *mdpi.comL WangSensors, 2017•mdpi.com*, Accessed: Jun. 09, 2024. [Online]. Available: https://www.mdpi.com/1424-8220/17/7/1572

[2] S. Palmal, S. Saha, N. Arya, and S. Tripathy, "CAGCL: Predicting Short- and Long-Time Breast Cancer Survival With Cross-Modal Attention and Graph Contrastive Learning," *IEEE J. Biomed. Heal. Informatics*, vol. 28, no. 12, pp. 7382–7391, Dec. 2024, doi: 10.1109/JBHI.2024.3449756.

[3] S. W. Duffy *et al.*, "The impact of organized mammography service screening on breast carcinoma mortality in seven Swedish counties: a collaborative evaluation," *Wiley Online Libr. Duffy, L Tabár, HH Chen, M Holmqvist, MF Yen, S Abdsalah, B Epstein, E FrodisCancer Interdiscip. Int. J. Am. 2002•Wiley Online Libr.*, vol. 95, no. 3, pp. 458–469, Aug. 2002, doi: 10.1002/cncr.10765.

[4] W. P. Evans, "Breast Cancer Screening: Successes and Challenges", doi: 10.3322/caac.20137.

[5] A. S. Majid, E. S. De Paredes, R. D. Doherty, N. R. Sharma, and X. Salvador, "Missed Breast Carcinoma: Pitfalls and Pearls," *Radiographics*, vol. 23, no. 4, pp. 881–895, 2003, doi: 10.1148/RG.234025083.

[6] A. Khan *et al.*, "A Recent Survey of Vision Transformers for Medical Image Segmentation," Dec. 2023, [Online]. Available: http://arxiv.org/abs/2312.00634

[7] H. Tanaka, S. Chiu, ... T. W.-P. in M., and undefined 2019, "Computer-aided diagnosis system for breast ultrasound images using deep learning," *iopscience.iop.orgH Tanaka, SW Chiu, T Watanabe, S Kaoku, T YamaguchiPhysics Med. Biol. 2019•iopscience.iop.org*, Accessed: Jun. 09, 2024. [Online]. Available: https://iopscience.iop.org/article/10.1088/1361-6560/ab5093/meta

[8] B. Gheflati, ... H. R.-C. of the I. E. in, and undefined 2022, "Vision transformers for classification of breast ultrasound images," *ieeexplore.ieee.orgB Gheflati, H Rivaz2022 44th Annu. Int. Conf. IEEE Eng. 2022•ieeexplore.ieee.org*, Accessed: Jun. 09, 2024. [Online]. Available: https://ieeexplore.ieee.org/abstract/document/9871809/

[9] J. An *et al.*, "Transformer-Based Weakly Supervised Learning for Whole Slide Lung Cancer Image Classification," *IEEE J. Biomed. Heal. Informatics*, vol. 10, no. 10, pp. 1–14, 2024, doi: 10.1109/JBHI.2024.3425434.

[10] W. Jia, M. Sun, J. Lian, and S. Hou, "Feature dimensionality reduction: a review," *Complex Intell. Syst.*, vol. 8, no. 3, pp. 2663–2693, Jun. 2022, doi: 10.1007/s40747-021-00637-x.

[11] V. Berisha *et al.*, "Digital medicine and the curse of dimensionality," *npj Digit. Med.*, vol. 4, no. 1, p. 153, Oct. 2021, doi: 10.1038/s41746-021-00521-5.

[12] A. Khan, S. H. Khan, M. Saif, A. Batool, A. Sohail, and M. Waleed Khan, "A Survey of Deep Learning Techniques for the Analysis of COVID-19 and their usability for Detecting Omicron," *J. Exp. Theor. Artif. Intell.*, pp. 1–43, Jan. 2023, doi: 10.1080/0952813X.2023.2165724.

[13] U. Ahmed, A. Khan, S. H. Khan, A. Basit, I. U. Haq, and Y. S. Lee, "Transfer learning and meta classification based deep churn prediction system for telecom industry," 2019.

[14] S. H. Khan, "Malaria parasitic detection using a new deep boosted and ensemble learning framework," *arXiv Prepr. arXiv2212.02477*, 2022.

[15] S. H. Khan, A. Sohail, A. Khan, and Y.-S. Lee, "COVID-19 Detection in Chest X-ray Images Using a New Channel Boosted CNN," *Diagnostics*, vol. 12, no. 2, p. 267, Jan. 2022, doi: 10.3390/diagnostics12020267.

[16] M. M. Zafar *et al.*, "Detection of tumour infiltrating lymphocytes in CD3 and CD8 stained histopathological images using a two-phase deep CNN," *Photodiagnosis Photodyn. Ther.*, vol. 37, p. 102676, Mar. 2022, doi: 10.1016/j.pdpdt.2021.102676.

[17] S. H. Khan *et al.*, "COVID-19 detection and analysis from lung CT images using novel channel boosted CNNs," *Expert Syst. Appl.*, vol. 229, p. 120477, Nov. 2023, doi: 10.1016/j.eswa.2023.120477.

[18] S. H. Khan, N. S. Shah, R. Nuzhat, A. Majid, H. Alquhayz, and A. Khan, "Malaria parasite classification framework using a novel channel squeezed and boosted CNN," *Microscopy*, May 2022, doi: 10.1093/jmicro/dfac027.

[19] Z. Rauf, A. Sohail, S. H. Khan, A. Khan, J. Gwak, and M. Maqbool, "Attention-guided multi-scale deep object detection framework for lymphocyte analysis in IHC histological images," *Microscopy*, vol. 72, no. 1, pp. 27–42, 2023.

[20] S. H. Khan, R. Iqbal, and S. Naz, "A Recent Survey of the Advancements in Deep Learning Techniques for Monkeypox Disease Detection," *arXiv Prepr. arXiv2311.10754*, 2023.

[21] E. Kalafi, A. Jodeiri, S. Setarehdan, N. Lin, K. R.- Diagnostics, and undefined 2021, "Classification of breast cancer lesions in ultrasound images by using attention layer and loss ensemble in deep convolutional neural networks," *mdpi.com*, Accessed: Jun. 09, 2024. [Online]. Available: https://www.mdpi.com/2075-4418/11/10/1859

[22] Y. Luo, Q. Huang, X. L.-P. Recognition, and undefined 2022, "Segmentation information with attention integration for classification of breast tumor in ultrasound image," *ElsevierY Luo, Q Huang, X LiPattern Recognition, 2022•Elsevier*, doi: 10.1016/j.patcog.2021.108427.

[23] M. B.-B. S. P. and Control and undefined 2021, "Breast mass classification with transfer learning based on scaling of deep representations," *ElsevierM ByraBiomedical Signal Process. Control. 2021•Elsevier*, Accessed: Jun. 09, 2024. [Online]. Available: https://www.sciencedirect.com/science/article/pii/S1746809421004250

[24] R. Du *et al.*, "Discrimination of breast cancer based on ultrasound images and convolutional neural network," *iopscience.iop.orgR Du, Y Chen, T Li, L Shi, Z Fei, Y LiJournal Oncol. 2022•Wiley Online Libr.*, vol. 2022, 2022, doi: 10.1155/2022/7733583.

[25] S. Han, H. Kang, J. Jeong, ... M. P.-P. in M., and undefined 2017, "A deep learning framework for supporting the classification of breast lesions in ultrasound images," *iopscience.iop.orgS Han, HK Kang, JY Jeong, MH Park, W Kim, WC Bang, YK SeongPhysics Med. Biol. 2017•iopscience.iop.org*, Accessed: Jun. 09, 2024. [Online]. Available: https://iopscience.iop.org/article/10.1088/1361-6560/aa82ec/meta

[26] X. Qi *et al.*, "Automated diagnosis of breast ultrasonography images using deep neural networks," *ElsevierX Qi, L Zhang, Y Chen, Y Pi, Y Chen, Q Lv, Z YiMedical image Anal. 2019•Elsevier*, Accessed: Jun. 09, 2024. [Online]. Available: https://www.sciencedirect.com/science/article/pii/S1361841518306832

[27] M. Hassanien, V. Singh, D. Puig, M. A.-N.- Diagnostics, and undefined 2022, "Predicting breast tumor malignancy using deep ConvNeXt radiomics and quality-based score pooling in ultrasound sequences," *mdpi.comMA Hassanien, VK Singh, D Puig, M Abdel-NasserDiagnostics, 2022•mdpi.com*, Accessed: Jun. 09, 2024. [Online]. Available: https://www.mdpi.com/2075-4418/12/5/1053

[28] W. Moon, Y. Lee, H. Ke, S. Lee, ... C. H.-C. methods and, and undefined 2020, "Computer-aided diagnosis of breast ultrasound images using ensemble learning from convolutional neural networks," *ElsevierWK Moon, YW Lee, HH Ke, SH Lee, CS Huang, RF Chang. methods programs Biomed. 2020•Elsevier*, Accessed: Jun. 09, 2024. [Online]. Available: https://www.sciencedirect.com/science/article/pii/S0169260719307059

[29] E. Evain, C. Raynaud, C. Ciofolo-Veit, ... A. P.-D. and, and undefined 2021, "Breast nodule classification with two-dimensional ultrasound using Mask-RCNN ensemble aggregation," *ElsevierE Evain, C Raynaud, C Ciofolo-Veit, A Popoff, T Caramel. P Kbaier, C BalleyguierDiagnostic Interv. Imaging, 2021•Elsevier*, Accessed: Jun. 09, 2024. [Online]. Available: https://www.sciencedirect.com/science/article/pii/S2211568421002X

[30] M. Ragab, A. Albukhari, J. Alyami, R. M.- Biology, and undefined 2022, "Ensemble deep-learning-enabled clinical decision support system for breast cancer diagnosis and classification on ultrasound images," *mdpi.comM Ragab, A Albukhari, J Alyami, RF MansourBiology, 2022•mdpi.com*, Accessed: Jun. 09, 2024. [Online]. Available: https://www.mdpi.com/2079-7737/11/3/439

[31] S. Misra *et al.*, "Ensemble Transfer Learning of Elastography and B-mode Breast Ultrasound Images," Feb. 2021, [Online]. Available: http://arxiv.org/abs/2102.08567

[32] Y. Yang, T. Xiang, X. Lv, L. Li, L. M. Lui, and T. Zeng, "Double Transformer Super-Resolution for Breast Cancer ADC Images,"

*IEEE J. Biomed. Heal. Informatics*, vol. 28, no. 2, pp. 917–928, Feb. 2024, doi: 10.1109/JBHI.2023.3341250.

[33] D. Hendrycks and K. Gimpel, "Gaussian Error Linear Units (GELUs)," Jun. 2016, Accessed: Mar. 30, 2024. [Online]. Available: http://arxiv.org/abs/1606.08415

[34] M. Kaur, D. Singh, A. A. Alzubi, A. Shankar, and U. Rawat, "DARNet: Deep Attention Module and Residual Block-Based Lung and Colon Cancer Diagnosis Network," *IEEE J. Biomed. Heal. Informatics*, vol. 3, no. 10, pp. 1–8, 2024, doi: 10.1109/JBHI.2024.3502636.

[35] M. Yap, G. Pons, J. Marti, … S. G.-I. journal of, and undefined 2017, "Automated breast ultrasound lesions detection using convolutional neural networks," *ieeexplore.ieee.orgMH Yap, G Pons, J Marti, S Ganau, M Sentis, R Zwiggelaar, AK Davison, R MartíIEEE J. Biomed. Heal. informatics, 2017•ieeexplore.ieee.org*, Accessed: Jun. 09, 2024. [Online]. Available: https://ieeexplore.ieee.org/abstract/document/8003418/

[36] S. H. Khan *et al.*, "COVID-19 detection in chest X-ray images using deep boosted hybrid learning," 2021. doi: 10.1016/j.compbiomed.2021.104816.

[37] J. Wang *et al.*, "Information bottleneck-based interpretable multitask network for breast cancer classification and segmentation," *ElsevierJ Wang, Y Zheng, J Ma, X Li, C Wang, J Gee, H Wang, W HuangMedical Image Anal. 2023•Elsevier*, Accessed: Jun. 20, 2024. [Online]. Available: https://www.sciencedirect.com/science/article/pii/S1361841522003152

[38] A. Dosovitskiy *et al.*, "An Image is Worth 16x16 Words: Transformers for Image Recognition at Scale," Oct. 2020, doi: org/10.48550/arXiv.2010.11929.

[39] A. Iqbal and M. Sharif, "BTS-ST: Swin transformer network for segmentation and classification of multimodality breast cancer images," *Knowledge-Based Syst.*, vol. 267, p. 110393, May 2023, doi: 10.1016/J.KNOSYS.2023.110393.

[40] B. Shareef, M. Xian, A. Vakanski, and H. Wang, "Breast Ultrasound Tumor Classification Using a Hybrid Multitask CNN-Transformer Network," 2023, pp. 344–353. doi: 10.1007/978-3-031-43901-8_33.

[41] B. Xu, K. Wu, Y. Wu, J. He, and C. Chen, "Dynamic adversarial domain adaptation based on multikernel maximum mean discrepancy for breast ultrasound image classification," *Expert Syst. Appl.*, vol. 207, p. 117978, Nov. 2022, doi: 10.1016/j.eswa.2022.117978.